\documentclass[fleqn,usenatbib]{mnras}

\usepackage{times}

\usepackage[T1]{fontenc}
\usepackage{ae,aecompl}

\usepackage{amsmath}
\usepackage{graphicx}
\usepackage{lineno}

\newcommand{\pluseq}{\mathrel{+}=}
\newcommand{\diff}{\mathrm{d}}

\title[Efficient Two-Point Correlations]{A Computationally Efficient Approach
for Calculating Galaxy Two-Point Correlations}

\author[R. Demina et al.]{
Regina Demina,$^{1}$\thanks{E-mail: regina@pas.rochester.edu},
Sanha Cheong,$^{1}$\thanks{E-mail: scheong@u.rochester.edu},
Segev BenZvi,$^{1}$\thanks{E-mail: sybenzvi@pas.rochester.edu}
Otto Hindrichs,$^{1}$\thanks{Email: otto.heinz.hindrichs@cern.ch}
\\
$^{1}$Department of Physics and Astronomy, University of Rochester,
500 Joseph C. Wilson Boulevard, Rochester, NY 14627, USA
}

\date{Accepted XXX. Received YYY; in original form ZZZ}

\pubyear{2016}

\begin{document}
\label{firstpage}
\pagerange{\pageref{firstpage}--\pageref{lastpage}}
\maketitle

%
%

\begin{abstract}

We developed a modification to the calculation of the two-point correlation
function commonly used in the analysis of large scale structure in cosmology.
An estimator of the two-point correlation function is constructed by
contrasting the observed distribution of galaxies with that of a uniformly
populated random catalog.  Using the assumption that the distribution of random
galaxies in redshift is independent of angular position allows us to replace
pairwise combinatorics with fast integration over probability maps.  The new
method significantly reduces the computation time while simultaneously
increasing the precision of the calculation. It also allows to introduce cosmological parameters only
at the last and least computationally expensive stage, which is helpful when exploring 
various choices for these parameters.

\end{abstract}



\section{Introduction}\label{sec:intro}
The two-point spatial correlation function (2pcf) is a key tool used in
astrophysics for the study of large scale structure (LSS). Given an object such as a
galaxy, the number of other galaxies within a volume element
$\Delta V$ at a distance $s$ from the first galaxy is
\begin{equation}
  \Delta N = \overline{n}(1 + \xi(s))\Delta V,
\end{equation}
where $\overline{n}$ is the mean number density and $\xi(s)$ is the
two-point correlation function characterizing the deviation from the uniform distribution in separation between galaxies
\citep{Peebles:1973}.

The correlation function $\xi(s)$ and its Fourier transform, the galaxy power
spectrum $\mathcal{P}(k)$, have been used 
to describe the distribution of matter in galaxy surveys
\citep{Yu:1969,Peebles:1974,Groth:1977gj,Feldman:1993}. During the past decade
$\xi(s)$ has become a popular tool for the reconstruction of the clustering
signal known as baryon acoustic oscillations, or BAO \citep{Eisenstein:2005su}.
The BAO signal is the signature of the density differences that arose in the
early universe before the thermal decoupling of photons and baryons
\citep{Sunyaev:1970eu,Peebles:1970ag}. It is detectable today as a
characteristic peak in the galaxy spatial correlation function at roughly
$s=110h^{-1}$~Mpc, where $h$ defines the Hubble parameter
$H_0=100h$~km~s$^{-1}$~Mpc$^{-1}$.

Given a spectroscopic survey containing the 3D coordinates of each galaxy, there
exist several possible ways to estimate $\xi(s)$. The most popular estimator is
due to \cite{Landy:1993yu}, which is
constructed by combining pairs of galaxies from a catalog of observed objects
$D$ (``data'') and a randomly generated catalog $R$ with galaxies
distributed uniformly over the fiducial volume of the survey but using the same
selection function as the data. The Landy-Szalay (LS) estimator is
\begin{equation}\label{eq:2pcf}
  \hat{\xi}(s) = \frac{DD(s) - 2DR(s) + RR(s)}{RR(s)},
\end{equation}
where $DD$, $RR$, and $DR$ are the normalized distributions of the pairwise
combinations of galaxies from the data and random catalogs (plus cross terms)
at a given distance $s$ from each other.

There exist many 2pcf estimators other than the LS estimator, and they have different advantages and limitations. However, all commonly known estimators are functions of $DR$ and/or $RR$ distributions (see \cite{Hamilton:1993fp}; \cite{Kerscher:1999hc}; \cite{VargasMagana:2012nu}) and therefore depend on the random catalog $R$. To limit statistical fluctuations in $\hat{\xi}(s)$, it is typical to generate
random catalogs with one to two orders of magnitude more galaxies than the
survey under investigation. Unfortunately, the ``brute-force'' computation of
$\hat{\xi}(s)$, in which all possible pair combinations are counted, is an
$\mathcal{O}(N_R^2)$ calculation due to $RR$, where $N_R$ is the number of galaxies in the random catalog $R$; this results in a trade-off between statistical
uncertainties and computation time. This trade-off has consequences beyond
reducing uncertainties. For example, researchers often simulate a variety of
cosmological parameters when studying the LSS, but the computational overhead
required to calculate $\hat{\xi}(s)$ may limit the number and type of possible
analyses that can be carried out. The overhead also increases the effort needed
to compute the covariance of the 2pcf as well as the effect of different sources of systematic uncertainties. 

In this paper, we suggest a method which substantially reduces the time needed
to compute $RR$ and $DR$ and hence is applicable to any estimator $\hat{\xi(s)}$\footnote{The code can be downloaded from {\url{http://www.pas.rochester.edu/~regina/LaSSPIA.html}}}.
Moreover, the computationally intensive part of the calculation is independent 
of the choice of cosmological parameters, allowing exploring a larger parameter space. 
The method is based on the assumption that the
probability for a galaxy to be observed at a particular location in the random
catalog can be factorized into separate angular and redshift components.
This assumption is frequently made in the analysis of large scale structures and it allows us to
replace pairwise combinatorics in the calculations of $RR$ and $DR$ with fast integration
over probability maps. We find that the estimation of $\hat{\xi}(s)$ significantly speeds up
with respect to the brute-force calculation. In practical
terms, this means an estimate of $\hat{\xi}(s)$ which is typically carried out
on large computing clusters can be performed on a modern notebook computer.

The paper is structured as follows. In Section~\ref{sec:proof} we describe the
mathematical justification of the method.
In Section~\ref{sec:algo} we describe the
resulting algorithm used to compute $\hat{\xi}(s)$ in detail. The performance
of the algorithm is studied in Section~\ref{sec:performance} using mock
catalogs and spectroscopic data from the CMASS portion of the Sloan Digital Sky
Survey (SDSS) DR9 catalog \citep{ Ross:2012qm, Sanchez:2012sg, Anderson:2012sa,
Percival:2013sga}.  We then conclude in Section~\ref{sec:conclusion}.

\section{Mathematical proof} \label{sec:proof}
\subsection{Random-random distribution}
The 3D position of any galaxy $\vec{r}$ is described by its right ascension $\alpha$, declination $\delta$,  and its redshift $z$.  Based on this information the cosmological distances are calculated given a set of cosmological parameters: $\Omega_M$ - the present day relative matter density of the universe,
$\Omega_k$ - the measure of the curvature of space, and $\Omega_\Lambda$ - the
relative density due to the cosmological constant. The comoving radial distance $r(z)$ is calculated from the observed redshift $z$ as
\begin{linenomath}
  \begin{equation} \label{eq:comovingr}
    r(z) = D_H I(z),\\
  \end{equation}
\end{linenomath}
    where
\begin{linenomath}
  \begin{equation} \label{eq:Hubble}
    D_H=c/H_0\\
  \end{equation}
\end{linenomath}
    is the Hubble distance,  $c$ is the speed of light and $I(z)$ is calculated as:
\begin{linenomath}
  \begin{equation} \label{eq:cosmoInt}
    I(z) = \int_0^z \diff z' \,
    \left(\Omega_M(1+z')^3+\Omega_k(1+z')^2+\Omega_\Lambda\right)^{-1/2}.
  \end{equation}
\end{linenomath}
The transverse distance $t(z)$ is calculated as:
\begin{linenomath}
  \begin{equation} 
  \label{eq:comovtrans}
    t(z) = \begin{cases} 
    D_H/\sqrt{\Omega_k}\ \sinh\big(\sqrt{\Omega_k}I(z)\big), \text{for }\Omega_k>0 \\
    r(z), \text{for }\Omega_k=0 \\ 
    D_H/\sqrt{|\Omega_k|}\ \sin\big(\sqrt{|\Omega_k|}I(z)\big), \text{for }\Omega_k<0 \\

    \end{cases}
  \end{equation}
\end{linenomath}

As suggested by current cosmological constraints (see e.g. \cite{Aubourg:2014, Ade:2015xua}), $\Omega_k$ is small and thus eq.~\ref{eq:comovtrans} can be approximated by
\begin{linenomath}
  \begin{equation} 
  \label{eq:comovtransapprox}
    t(z) = r(z) \Big( 1+\frac{\Omega_k}{6} \big(I(z)\big)^2\Big).\\
  \end{equation}
\end{linenomath}

%

\begin{figure}
  \begin{center}
  \vspace{-1.5cm}
    \includegraphics[width=1.0\linewidth]{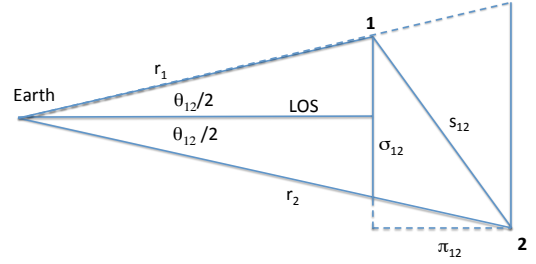}
  \end{center}
  \vspace{-1.5cm}
  \caption{Calculation of the distance between  points $1$ and $2$ given their comoving distances from the Earth $r_1$ and $r_2$ and their angular separation $\theta_{12}$. }
  \label{fig:geometry}
\end{figure}

 
The angular separation $\theta_{12}$ between  points $1$ and $2$  given their right ascension $\alpha$, declination $\delta$ is calculated as 
\begin{equation}\label{eq:costheta}
  \theta_{12} = \cos^{-1} \big(\cos{\delta_1}\cos{\delta_2}
  \cos{(\alpha_1-\alpha_2)}+\sin{\delta_1}\sin{\delta_2}\big).
\end{equation}
The distance $s_{12}$ between  these two points (illustrated in Fig.~\ref{fig:geometry}) is approximated by: 
\begin{equation}
\label{eq:scurved}
s_{12} = \sqrt{\sigma_{12}^2+\pi_{12}^2},
\end{equation}
where $\sigma_{12}$ and $\pi_{12}$, the distances transverse and parallel to the line of sight (LOS) respectively, are defined as:
\begin{gather}
\sigma_{12} = (t_1+t_2) \sin{\frac{\theta_{12}}{2}},\label{eq:sigma}	\\
\pi_{12} = \left|r_1-r_2\right|\cos{\frac{\theta_{12}}{2}}.
\label{eq:pi}
\end{gather}
\begin{figure}
  \begin{center}
  \vspace{-0.3cm}
  \hspace{-0.1cm}
    \includegraphics[width=1.03\linewidth]{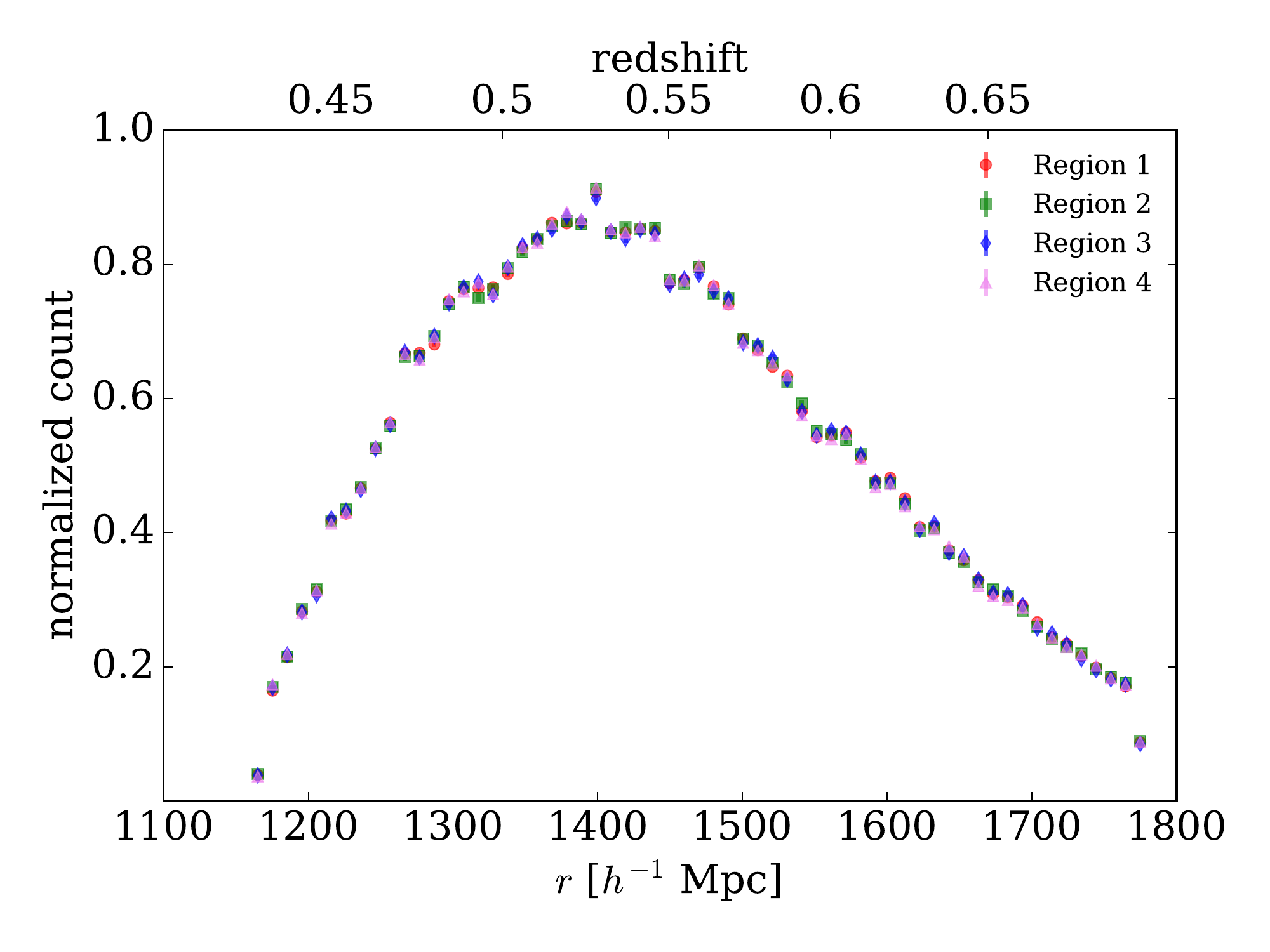}
  \end{center}
  \vspace{-0.5cm}
  \caption{Distribution of galaxies over redshift and comoving distance $r$, calculated using cosmological parameters defined in Section~\ref{sec:performance}, in four widely
  separated regions of the Northern sky based on the SDSS-III DR9 random catalog. Regions are defined by the following selection criteria: region 1 ($-4<\delta<8^o$, $125<\alpha<250^o$), region 2 ($8<\delta<57^o$, $108<\alpha<135^o$), region 3 ($25<\delta<50^o$, $135<\alpha<225^o$), region 4 ($8<\delta<40^o$, $225<\alpha<265^o$).}
  \label{fig:PrettyR}
\end{figure}
Since random catalogs are typically generated by uniformly populating the
fiducial volume of the survey, the galaxy distribution over $z$, and thus over $r$, is factorizable
from the angular distribution. In other words, any angular region of the sky
has the same distribution of galaxies in $z$ (see
\cite{Ross:2012qm} and Fig.~\ref{fig:PrettyR}). This means that the expected
count of random galaxies, $R(\vec{r})$, can be factorized into the product of the
expected count $R_\text{ang}(\alpha,\delta)$ at a given angular position and a
redshift probability density function (PDF) $P_z(z)$:
\begin{linenomath}
\begin{equation} \label{eq:factorized}
  R(\vec{r}) =
  R_\text{ang}(\alpha, \delta)
  P_z(z).
\end{equation}
\end{linenomath}
$R(\vec{r})$ evaluated this way has smaller statistical uncertainty since the precision of $R_\text{ang}(\alpha, \delta)$ depends on the
number of points in the 2D angular cell, and the precision of $P_z(z)$ depends on the number of points in the 1D range in $z$. 
In contrast, the statistical precision without the factorizability assumption is
determined by the number of points in a 3D cell $(z,\alpha,\delta)$.
We will show in Section~\ref{sec:performance} that the uncertainties in
$\hat{\xi}(s)$ evaluated using the suggested method are indeed smaller.

The expected count of random-random galaxy pairs separated by a distance
$s$ can be expressed as:
\begin{linenomath}
\begin{multline} \label{eq:RR1}
  RR(s) = \frac{1}{2}\int \diff \Omega_1 \diff z_1 \diff \Omega_2 \diff z_2 \,
  R_\text{ang}(\alpha_1, \delta_1)\, P_z(z_1) \\
  \hspace{1.2in} \times
  R_\text{ang}(\alpha_2, \delta_2)\, P_z(z_2)\,
  \delta(s-s_{12}), \\
\end{multline}
\end{linenomath}
where $\Omega$ represents a solid angle (and
$\diff\Omega=\cos{\delta}\, \diff\delta\, \diff\alpha$), $s_{12}$ is calculated according to eq.~\ref{eq:scurved}.
The integral is taken over the entire fiducial volume of the random
catalog $R$, and a factor of $1/2$ is introduced to account for double counting of the random-random pairs.
The Dirac-$\delta$ function is introduced to ensure that the
distance between two galaxies $s_{12}$ is equal to the distance of interest
$s$. To isolate the angular variables, we rewrite the $\delta$ function
in eq.~\ref{eq:RR1} as an integral of the product of two $\delta$ functions:
\begin{linenomath}
\begin{equation} \label{eq:deltafunction}
  \delta(s-s_{12}) = \int \diff\theta\,
  \delta\left( s-s_{12} \right)
  \delta(\theta - \theta_{12}).
\end{equation}
\end{linenomath}
Note that the second $\delta$ function is independent of the radial
positions $r_1,r_2$ of the galaxies, or their redshifts. Thus, $RR$ can be rewritten as
\begin{linenomath}
\begin{multline} \label{eq:RR2}
  RR(s) = \int \diff z_1 \diff z_2 \diff \theta \,  P_z(z_1)  P_z(z_2)\, f(\theta) \\
  \times
  \delta\left(s-s_{12}\right),
\end{multline}
\end{linenomath}
where
\begin{linenomath}
\begin{multline} \label{eq:ftheta}
  f(\theta) = \frac{1}{2} \int \diff \Omega_1  \diff \Omega_2\,
  R_\text{ang}(\alpha_1, \delta_1)\,
  R_\text{ang}(\alpha_2, \delta_2)\,	\\
  \times
  \delta(\theta-\theta_{12})
\end{multline}
\end{linenomath}
is the count of $RR$ galaxy pairs whose angular separation is $\theta$.

The count of random galaxy pairs is
constructed in two steps: 
\begin{enumerate}
\item  \begin{itshape} histogramming\end{itshape}, where we construct the distribution $f(\theta)$ over angular
separation
using the count of random galaxies $R_\text{ang}(\alpha, \delta)$ according to eq.~\ref{eq:ftheta}, 
\item \begin{itshape} integration\end{itshape},
where we convolve the angular distribution with the redshift
PDFs $P_z(z_1)$ and $P_z(z_2)$ to obtain the distribution of random pairs over $s$. 
\end{enumerate}
Note that only the second step depends on the choice of cosmological parameters. 
\subsection{Data-random and data-data distributions}
Unlike in the random catalog, in the data catalog the distribution of galaxies
over $z$ cannot be factorized from their angular distribution. Thus,
the count distribution over  the opening angle between
galaxies in the data and random catalogs is also a function of $z_1$, the
position of the data galaxy:
\begin{linenomath}
  \begin{multline} \label{eq:P2D}
    g(\theta,z_1) = 
    \int \diff\Omega_1  \diff\Omega_2 \, D(z_1, \alpha_1, \delta_1)\,
    R_\text{ang}(\alpha_2, \delta_2) \\
    \times
  \delta(\theta-\theta_{12}).
  \end{multline}
\end{linenomath}
The $DR$ distribution can then be calculated as
\begin{linenomath}
\begin{multline} \label{eq:DR2}
  DR(s) = \int  \diff z_1  \diff z_2 \diff\theta \, P_z(z_2)\, g(\theta,z_1) \\
  \times
  \delta\left(s-s_{12}\right).
\end{multline}
\end{linenomath}
%
At the histogramming step, we construct the distribution $g(\theta,z_1)$ 
according to eq.~\ref{eq:P2D}, keeping track of the redshift of the data galaxies $z_1$. At the integration step $g(\theta,z_1)$ is convolved with the redshift
PDF $P_z(z_2)$ to obtain the distribution of data-random pairs over $s$ according to eq.~\ref{eq:DR2}.

Finally, the data-data count $DD$ is estimated using a brute-force iteration over all possible data galaxy pairs. For uniformity, the $DD$ calculation is also broken down into histogramming and integration steps. At the histogramming step the count distribution over the opening angle $u(\theta,z_1, z_2)$ is constructed from the pair count of data galaxies keeping track of the redshifts of both galaxies in the pair.

At the integration step, $DD(s)$ is evaluated from $u(\theta, z_1, z_2)$ by converting the redshifts into distances and calculating the distance between the two galaxies $s$. This breakdown does not offer any time savings in the calculation of $DD$, but it allows for the definition of the cosmological parameters only at the second step. This way the computationally intensive first step can be performed once and then the computationally fast second step is performed for each set of cosmological parameters. 
\subsection{Generalization of the 2pcf to anisotropic case}
\label{sec:generalization}
The discussion up to this point has dealt with only a spherically symmetric 2pcf.
However, the algorithm is easily generalized to study anisotropy in the 2pcf \citep{Davis:1983apj}. 

The histogramming step is identical to the isotropic case.  
At the integration step, we compute $RR$, $DR$ and $DD$ in two dimensions $(\sigma,\pi)$:
\begin{linenomath}
\begin{multline} \label{eq:RRsigmapi}
  RR(\sigma,\pi) = \frac{1}{N_{RR}}\int  \diff z_1  \diff z_2 d\theta \, P_z(z_1)\, P_z(z_2)\,
  f(\theta)  \\
  \times
  \delta(\sigma-\sigma_{12})\delta(\pi-\pi_{12}),
\end{multline}
\end{linenomath}
\begin{linenomath}
\begin{multline} \label{eq:DRsigmapi}
  DR(\sigma,\pi) = \frac{1}{N_{DR}}\int  \diff z_1  \diff z_2 \diff \theta \, P_z(z_2)\,
  g(\theta,z_1) \\
  \times
  \delta(\sigma-\sigma_{12})\delta(\pi-\pi_{12}),
\end{multline}
\end{linenomath}
where  the distances transverse ($\sigma$) and parallel ($\pi$) to the LOS are computed according to 
eqs.~\ref{eq:sigma} and \ref{eq:pi} respectively.  

 The BAO signal is expected to manifest itself 
 as an ellipse in this 2D histogram. In the isotropic case the ellipse is reduced to a circle. 
\section{Description of the algorithm}
\label{sec:algo}
\subsection{Weights}
To data we apply the weights according to the prescription of \cite{Ross:2012qm} to deal with the issues of close-pair corrections ($w_\text{cp}$), redshift-failure corrections ($w_\text{rf}$), systematic targeting effects ($w_\text{sys}$) and shot noise and cosmic variance ($w_\text{FKP}$) \citep{Feldman:1993} such that the total weight of a given data galaxy is:
\begin{equation}
\label{eq:dataweights}
 w^D = w_{\text{FKP}}\cdot w_{\text{sys}}\cdot(w_{\text{rf}}+w_{\text{cp}}-1).
\end{equation}
Each galaxy in the random catalog has a $z$-dependent
weight $w^R(z)$, defined the same way as $w_\text{FKP}$ in data catalog. The algorithm can be generalized to
the case where the random weights also depend on angular position and are factorizable
into angular and redshift weights: $w^R=w^R(z)w^R(\alpha,\delta)$.
\subsection{Binning}
$RR$ and $DR$ are calculated using finely binned probability densities in $(\alpha,\delta)$ and $z$ defined from the existing random catalog, or based on the completeness map and the radial selection function. The choice of the bin sizes is
important and is determined by the final bin size $\Delta s$ desired in
$\hat{\xi}(s)$. In practice, this means that the bin sizes in $\alpha$,
$\delta$ and $\theta$ must be smaller than the angle $\theta_{min}$ subtended by $\Delta s$ at the
outermost radius of the data set, $R_{max}$ at least by a factor of two ($\theta_{min}=\Delta s /2/R_{max}$). The bin size in $z$ should be chosen such that the corresponding $\Delta r(z)$ be smaller than $\Delta s$ by the same factor. We note that the binning is not equal-area but for fine enough bins it does not affect the result. Some assumption about the cosmological parameters must be made for the calculation of $R_{max}$ and $r(z)$, so to be on the conservative side one should use the largest $R_{max}$ for the set of the cosmological parameters under evaluation, which typically means using the smallest value of $\Omega_m$. 
Using this algorithm with steps smaller than 1 $h^{-1}$~Mpc is possible but impractical as 
it results in an angular map that is too finely segmented and thus very large and noisy. 
\subsection{Limits}
In the calculation presented in this paper, we only consider galaxy pairs separated by a
distance less than a certain maximum distance scale of interest $l_\text{max}$. For this, the 2D angular space is divided into regions of the size $(\Delta\alpha,\Delta\delta)$, such that $\Delta\alpha=\Delta\delta=l_\text{max}/r_{min}$, where $r_{min}=R_{min}\cos(\delta_{max})$.
Here, $R_{min}$ is the smallest radial distance of the survey for a set of cosmological parameters under consideration (typically corresponding to the largest value of $\Omega_m$), and $\delta_{max}$ is the maximum declination of the survey. 
The algorithm proceeds to calculate $f(\theta_{12})$, $g(\theta_{12}, z_1)$ and $u(\theta_{12}, z_1, z_2)$ only within one $(\Delta\alpha,\Delta\delta)$ region and its neighbors.
\subsection{Algorithm}
\subsubsection{Probability Density Functions}
Once the binning and limits are chosen, the algorithm proceeds as follows.  First, using the
random catalog, we produce the count distribution $R_\text{ang}(\alpha,\delta)$ and
PDF $P_z(z)$.  For each
galaxy, $P_z(z)$ is incremented by $w^R(z)/N_R$, and
$R_\text{ang}(\alpha,\delta)$ is incremented by 1. If the catalog contains the angular dependent weight  $w^R(\alpha,\delta)$ then $R_\text{ang}(\alpha,\delta)$ is incremented by that amount. 
%
\begin{figure}
  \begin{center}
    \includegraphics[width=\linewidth]{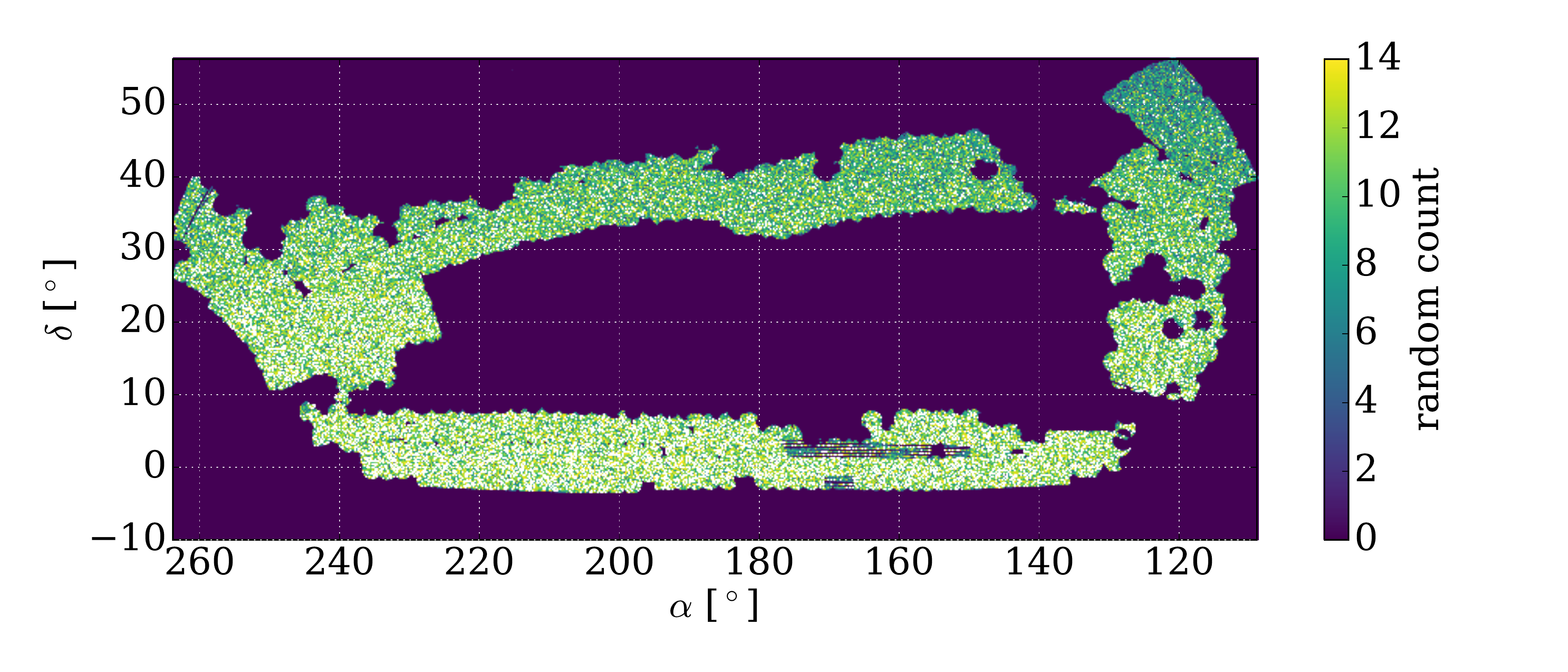}
    \includegraphics[width=\linewidth]{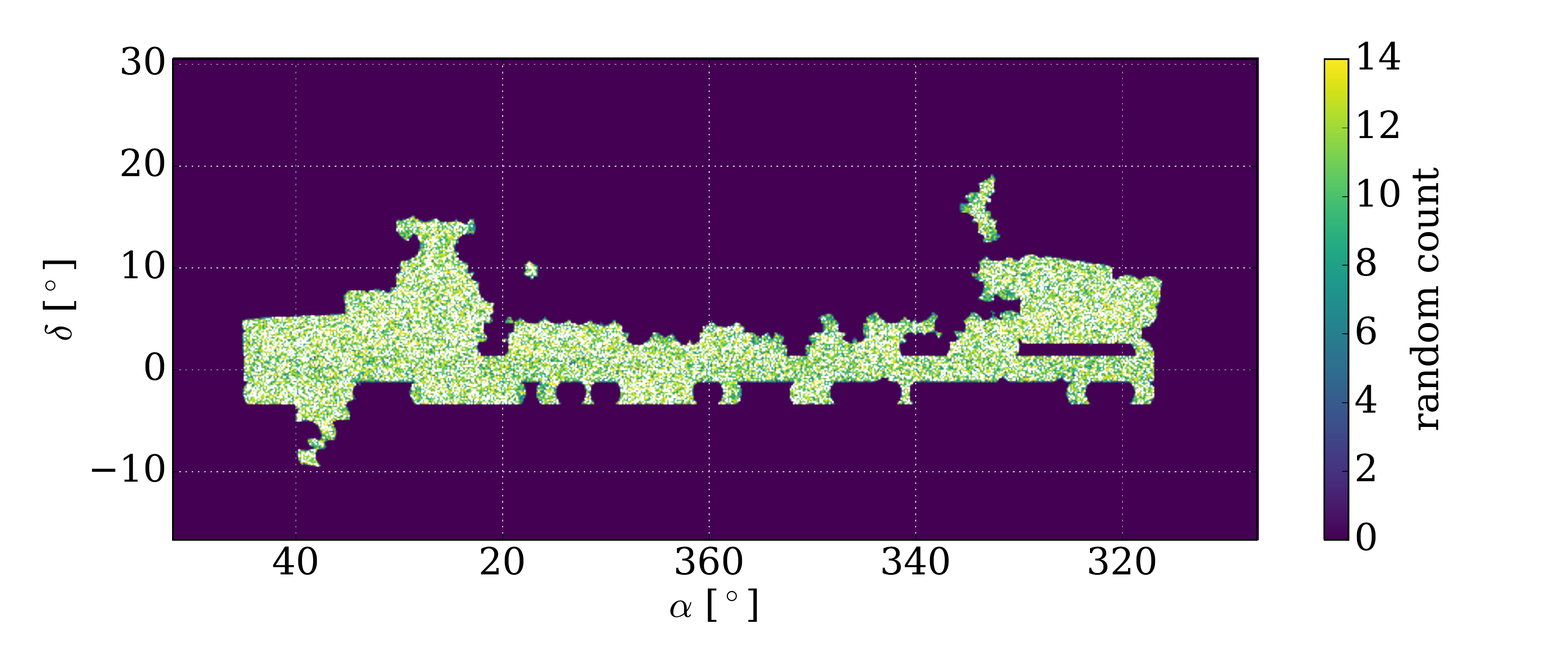}
    \includegraphics[width=0.95\linewidth]{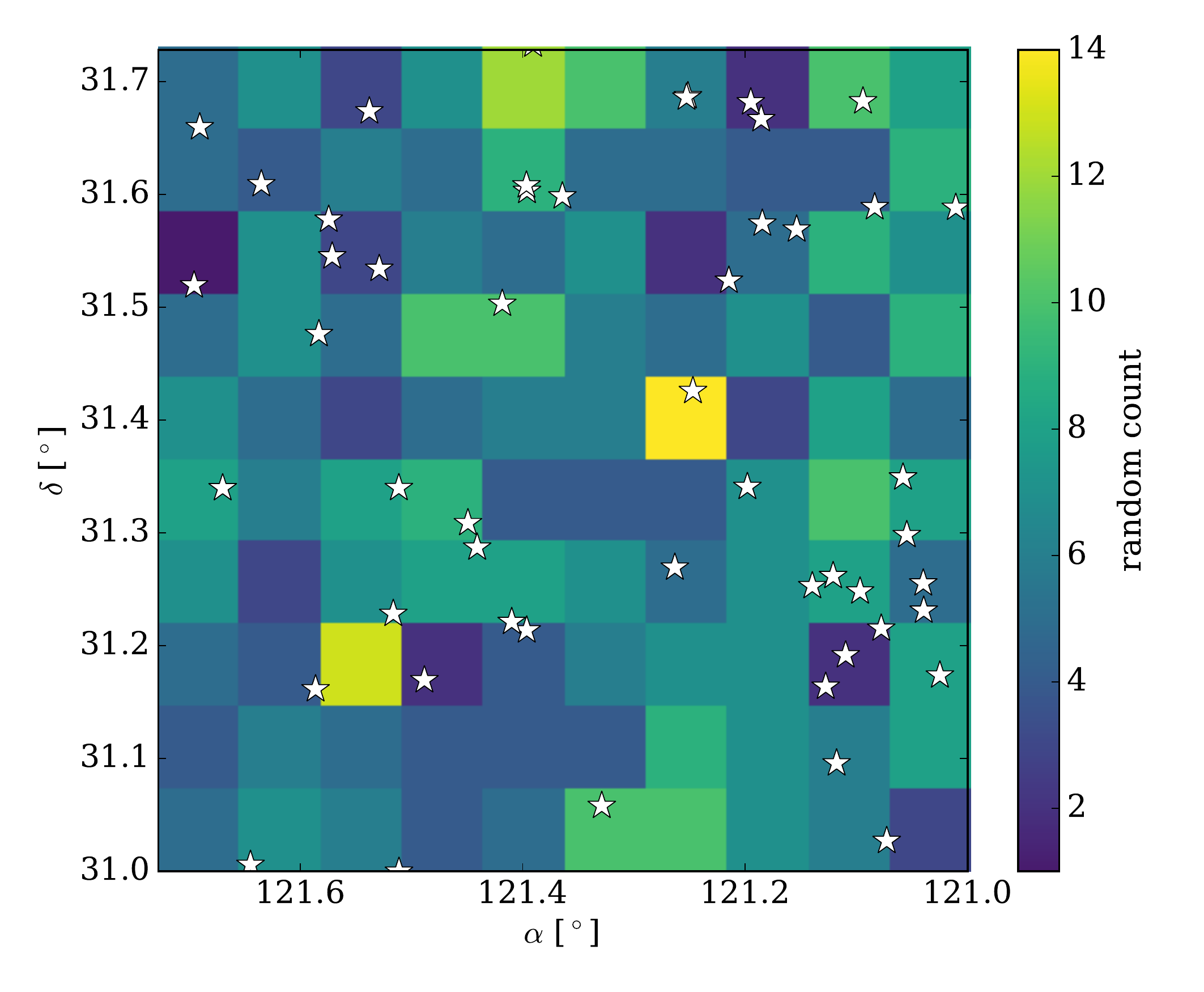}
  \end{center}
  \caption{{\sl Top}: Distribution of the DR9 random catalog in
    $(\alpha,\delta)$ in the northern sky.
    {\sl Middle}: the DR9 random catalog in the southern sky.
    {\sl Bottom}: a zoomed-in view of northern sky. The positions
    of galaxies observed in the DR9 survey are plotted as white stars.}
  \label{fig:rade}
\end{figure}
This part of the algorithm is linear in the number of galaxies in random catalog and hence is very fast. 

Examples of binned angular histograms are shown in
Fig.~\ref{fig:rade}. It is clear from the zoomed up view that the statistical fluctuations, determined by the finite size of the random catalog, are significant. However, it is important to note that these fluctuations are much smaller than the fluctuations in 3D cells with full coordinates $(z,\alpha,\delta)$,  effectively  used in the brute-force approach, or in any other approach that relies on 3D number density distribution. In this paper, we generate the 2D angular and 1D redshift probability maps based on an existing random catalog (published along with the observational data), but ideally these maps should be generated directly based on the tile completeness and radial selection function. This approach will create smoother probability maps and minimize the shot noise associated with the use of random catalogs.
\subsubsection{Histogramming}
 We compute the random-random 1D binned distribution $f(\theta_{12})$ using
$R_\text{ang}(\alpha,\delta)$.
To compute binned $f(\theta_{12})$, we consider all non-repeating pairs of angular cells $(\alpha_1,\delta_1)$ and $(\alpha_2,\delta_2)$. The angle $\theta_{12}$
between the two cells is calculated from their angular positions  using 
eq.~\ref{eq:costheta} and the corresponding bin is incremented: 
\begin{linenomath}
\begin{equation}
\label{eq:RRalgo}
f(\theta_{12}) \pluseq  a R_\text{ang}(\alpha_1,\delta_1) R_\text{ang}(\alpha_2,\delta_2),
\end{equation}
\end{linenomath}
where a factor is introduced such that $a=1/2$ when the two cells are identical  and $a=1$ otherwise.  

Then, we calculate the data-random 2D binned distribution $g(\theta_{12},z_1)$ using galaxies from the
data catalog and $R_\text{ang}(\alpha,\delta)$. Here, we consider every pair of a data galaxy $(z_1, \alpha_1, \delta_1)$ and an angular cell 
$(\alpha_2,\delta_2)$.  The angle $\theta_{12}$
between the data galaxy and the angular cell is calculated from their angular positions  using 
eq.~\ref{eq:costheta} and the corresponding 2D bin is incremented:
\begin{linenomath}
\begin{equation}
\label{eq:DRalgo}
 g(\theta_{12},z_1) \pluseq w^D R_\text{ang}(\alpha_2,\delta_2),
\end{equation}
\end{linenomath}
where $w^D$ is the weight of the data galaxy. 

We compute the data-data 3D binned distribution $u(\theta_{12}, z_1, z_2)$ by looping over all non-repeating pairs of galaxies. Each pair is weighted by the product of the weights of individual galaxies in the
survey:
\begin{linenomath}
\begin{equation}
\label{eq:DDalgo}
 u(\theta_{12}, z_1, z_2) \pluseq w^D_1\times w^D_2.
\end{equation}
\end{linenomath}
\subsubsection{Integration}
Next, we perform an integration over redshifts $z_1$ and $z_2$ and produce the
distributions $RR$, $DR$ and $DD$. This is achieved in three nested loops iterating
over $\theta_{12}$, $z_1$ and $z_2$. Given these three variables and choice of cosmological parameters, we compute the distance separation $s_{12}$ using eq.~\ref{eq:scurved} and increment the corresponding bin of the final distributions:
%
\begin{align}
  RR(s_{12}) &\pluseq f(\theta_{12})  P_z(z_1)  P_z(z_2), \label{eq:RRint} \\
  DR(s_{12}) &\pluseq  g(\theta_{12}, z_1)  P_z(z_2), \label{eq:DRint} \\
  DD(s_{12}) &\pluseq   u(\theta_{12}, z_1, z_2). \label{eq:DDint}
\end{align}
For the computation of the anisotropic 2pcf distances 
$\sigma_{12}$ and  $\pi_{12}$ are  computed according to eq.~\ref{eq:sigma} and eq.~\ref{eq:pi} respectively.
\subsubsection{Normalization}
The histograms are normalized in the following way.
For the unweighted calculation (all galaxies have weight $1$), the normalization constant of $RR$ is simply $N_{RR} = N_R(N_R-1)/2$.
However, for the weighted calculation, the normalization constant $N_{RR}$ is:
\begin{linenomath}
\begin{equation}
N_{RR} = \sum_{i=1}^{N_R}\sum_{j=i+1}^{N_R} w^R_i w^R_j
\end{equation}
\end{linenomath}
where $w^R_i$ is the weight of the $i^\text{th}$ galaxy in the random catalog. In this specific form, the calculation is $\mathcal{O}\left(N_R^2\right)$ and requires a double loop over indices. However, it can be rewritten as:
\begin{equation}
N_{RR} = \frac{1}{2} \left[\left(\sum_{i=1}^{N_R} w^R_i\right)^2 - \sum_{i=1}^{N_R} \left(w^R_i\right)^2\right]
\label{eq:N_RR}
\end{equation}
which is now an $\mathcal{O}\left(N_R\right)$ calculation. The same approach can be applied when normalizing the $DD$ histogram. 

For the unweighted calculation, the normalization constant of $DR$ is simply $N_{DR} = N_D\times N_R$ where $N_D$ is the total number of galaxies in the data catalog. For the weighted calculation, the normalization constant $N_{DR}$ is:
\begin{equation}\label{eq:N_DR}
N_{DR} =  \sum_{i=1}^{N_D} w^D_i\sum_{i=1}^{N_R}w^R_i
\end{equation}
where $w^D_i$ is the weight of the $i^\text{th}$ galaxy in the data catalog. The normalization constant $N_{DR}$ only requires an $\mathcal{O}\left(N_R\right)$ computation and hence can be computed quickly even in the weighted calculation.
Finally, the LS estimator is
calculated according to eq.~\ref{eq:2pcf}. All the other estimators can be calculated just as easily, based on $DD$, $RR$ and $DR$ distributions computed above.
\section{Performance of the algorithm}
\label{sec:performance}
\subsection{Algorithm settings and Random catalog generation}
The performance of the new algorithm is evaluated by calculating the runtime
and the uncertainty of the LS estimator of the 2pcf.
Though the method described in this paper can be applied under any cosmology, for certainty we assume a
$\Lambda$CDM+GR flat cosmology with parameters consistent with those used in
the analysis of the SDSS-III DR9 data set \citep{Anderson:2012sa}, i.e.
 $\Omega_M=0.274$ and $\Omega_\Lambda=0.726$. In the distance calculation we set $H_0=100h$~km~s$^{-1}$~Mpc$^{-1}$, and define the distances in units of $h^{-1}$~Mpc.
%
%

Usually, the calculations of $RR$ and
$DR$ are the most computationally expensive since the size of the random
catalog is typically much larger than the size of the data catalog.
Therefore, we report the results of $RR$ and $DR$ for different sizes of random
catalogs, while we do not change the data catalog.

To generate random catalogs of any desired size, we begin with the northern sky of the existing SDSS-III DR9 random catalog which contains $\sim$3.5M galaxies and extract the distributions
$R_\text{ang}(\alpha,\delta)$ and $P_z(z)$. We then randomly generate new
galaxies according to the product of these distributions. In each angular cell we generate 
the number of galaxies $N_R(\alpha,\delta)$, which is distributed according to Poisson distribution with a mean of 
$R_\text{ang}(\alpha,\delta)$. Each of these galaxies is assigned a redshift $z$, which is distributed according to $P_z(z)$.
Weights are a function of the redshift $z$. 
While the new
catalogs may amplify existing statistical fluctuations in the DR9 random
catalog, our interest is in testing the runtime of the algorithms and estimating the variation from the mean in $\hat{\xi}(s)$ as a function of random catalog size. We find the bias in the 2pcf, defined as deviations from the true mean (which is zero), is significantly smaller than statistical uncertainties (Fig.~\ref{fig:calib2}).
The bias is produced by shot noise in the random catalog, so if 
$R_\text{ang}(\alpha,\delta)$ and $P_z(z)$ are generated from the tile completeness and
radial selection function it can be minimized.  

%
\subsection{Timing study}
%
We compare the timing of the algorithm presented here against brute-force pair counting and the parallelized algorithm CUTE \citep{Alonso:2012arxiv}. In all cases, we report runtime in CPU hours rather than elapsed wall-clock time.
%
As in the fast algorithm, the brute-force pair count and computation with CUTE are performed using a grid scheme with a maximum distance of $ l_\text{max}=400h^{-1}$~Mpc. Hence, the three sets of results include the same amount of statistics.
%

\begin{figure}
\begin{center}
\includegraphics[width=\linewidth]{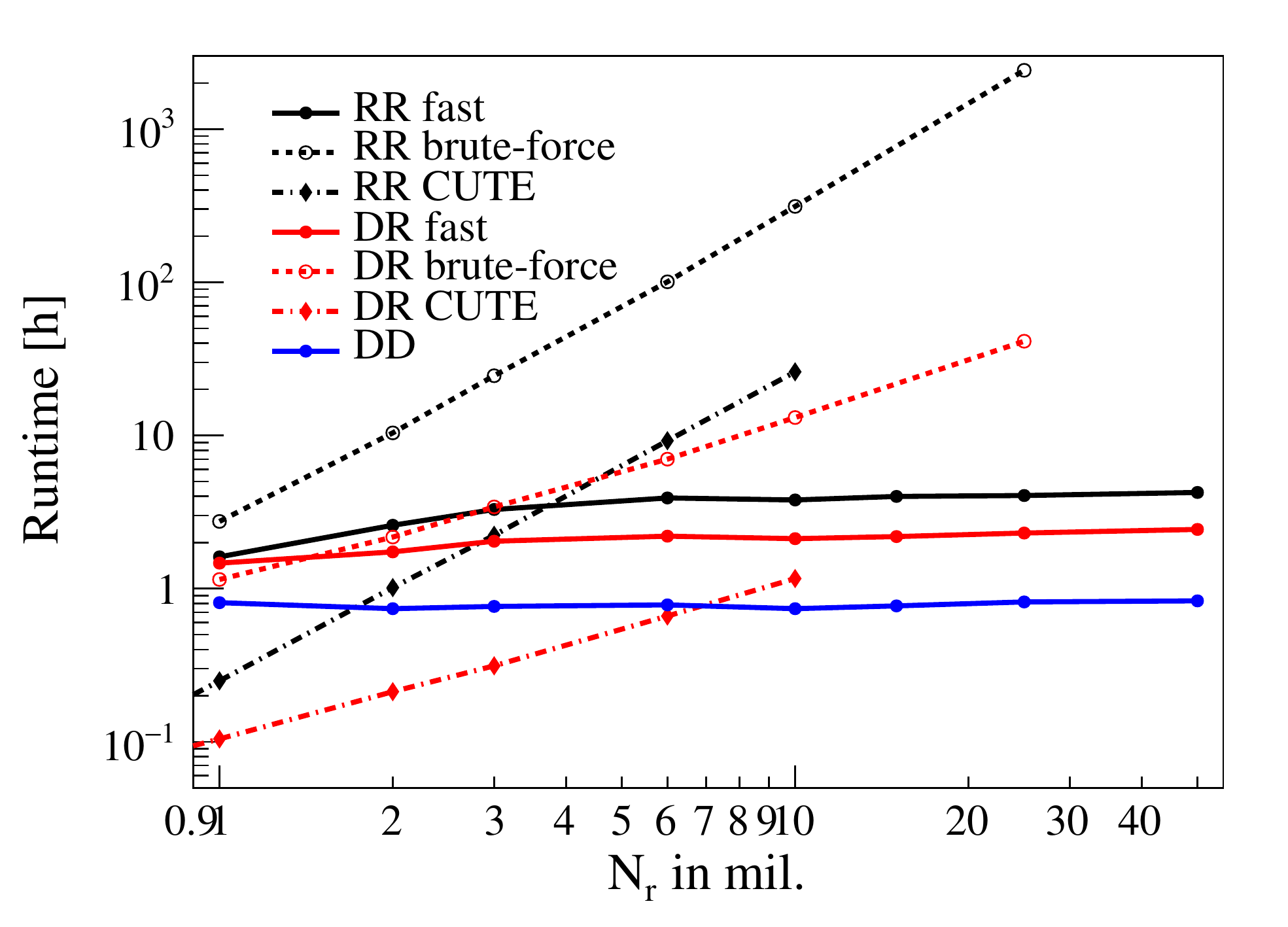}
\caption{The total runtime (in CPU hours) of the fast, brute-force and CUTE calculations of
$\hat{\xi}(s)$ as a function of the size of the random catalog.
\label{fig:runtime}}
\end{center}
\end{figure}

 In Fig.\,\ref{fig:runtime}, we show the runtime of the algorithms measured for
the pure algorithmic parts of the executable, which are based on a C++
implementation running on modern CPUs for random catalog sizes of $N_R=\text{1M}$ to 50M galaxies.
For this study we chose the following algorithm settings. 
The size of the  binning in the right ascension $\alpha$, declination $\delta$  and the angular separation $\theta$ is 1\,mrad, which corresponds  to the transverse
separation of 2.55\,$h^{-1}$~Mpc at the outermost radius. 
The binning in $r$ corresponds to a separation of 1\,$h^{-1}$~Mpc along the LOS.  
As expected, the runtime of our realization of the 
brute-force calculation as well as that of CUTE are proportional to $N_R^2$ for $RR$ and $N_R$ for
$DR$. In contrast, the runtime of the fast algorithm plateaus around a constant
value because it depends on the fiducial volume and the number of angular and radial bins, rather than
the size of the random catalog $N_R$. The maximum runtime of the fast algorithm is
reached when each bin in the $R_\text{ang}(\alpha, \delta)$ distribution is
populated and hence is used in the calculation of $\hat{\xi}(s)$.

\begin{figure}
\begin{center}
\includegraphics[width=\linewidth]{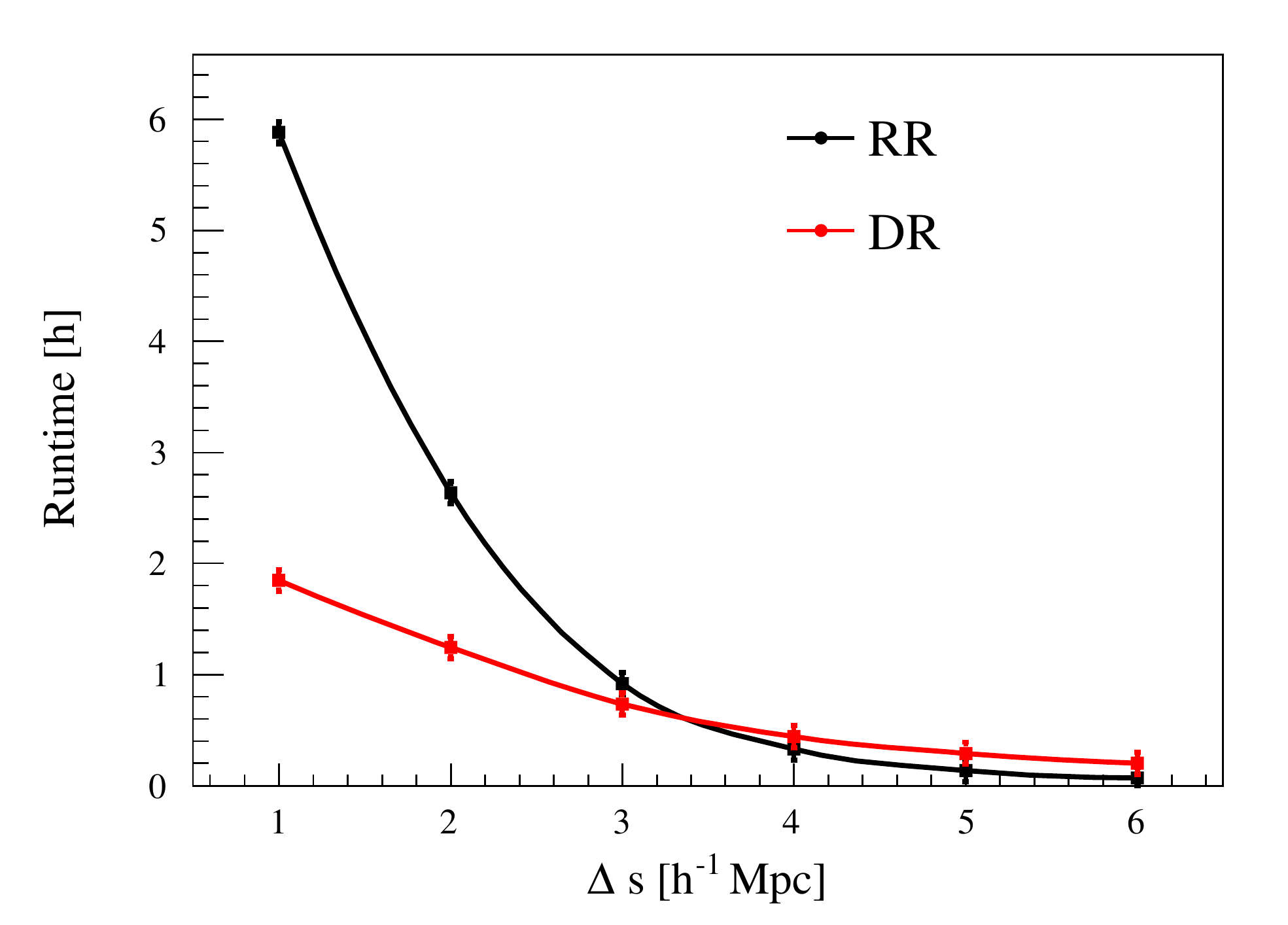}
\caption{The total runtime (in CPU hours) of the fast  calculation of
the pair-wise combinations as a function of the step in $\Delta s$. In this study 
$l_\text{max}=200\,h^{-1}$~Mpc. The binning in $r$ is $\Delta s/2$. The angular binning is $\Delta s/2/R_{max}$.  
\label{fig:timestep}}
\end{center}
\end{figure}

\begin{figure}
\begin{center}
\includegraphics[width=\linewidth]{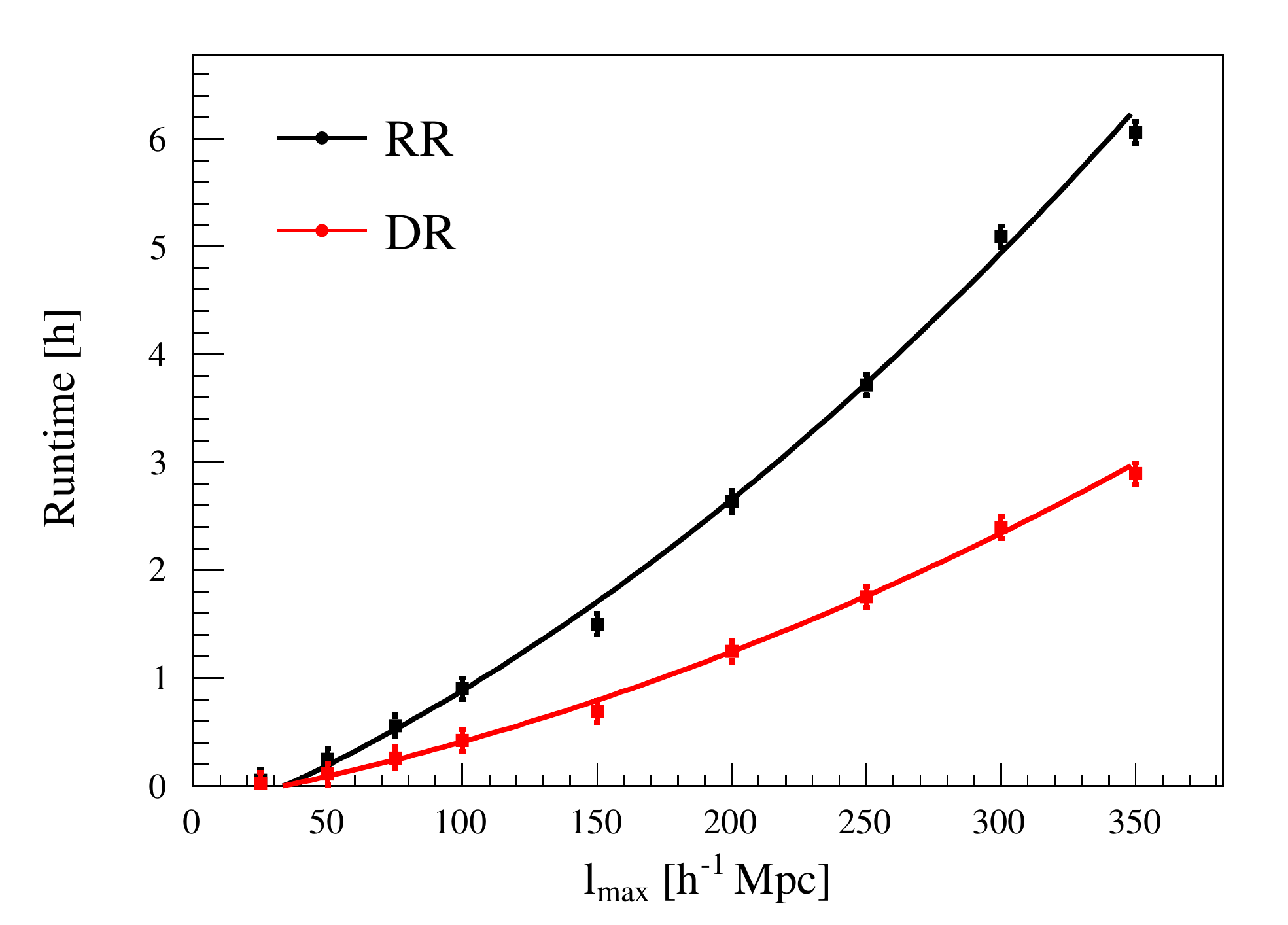}
\caption{The total runtime (in CPU hours) of the fast  calculation of
the pair-wise combinations as a function of the maximum distance $l_\text{max}$. The binning in $r$ is $1h^{-1}$~Mpc. The angular binning is 0.56\,mrad. 
\label{fig:timebox}}
\end{center}
\end{figure}

The fits of the runtime dependence on $\Delta s$, shown in Fig.~\ref{fig:timestep},  include  terms  proportional to $\Delta s^{-1}$ and $\Delta s^{-2}$. 
The fits of the runtime dependence on $l_\text{max}$, shown in Fig.~\ref{fig:timebox},  include terms linear and quadratic in $l_\text{max}$. The numbers of bins in $\alpha$, $N_{\alpha}$ and $\delta$, $N_{\delta}$ scale inversely with the bin
size $\Delta s$ in the final histogram, and linearly with the
distance scale of interest $l_\text{max}$. The total number of the 2D angular cells is the product of $N_{\alpha}$ and $N_{\delta}$. Thus, the calculation time scales linearly with the total number of the 2D angular cells. Other fast computational techniques based on the number density rather than the actual permutation
counting rely on data distribution in 3D cells \citep{Moore:2001} and thus have stronger scaling with the number of bins.
\subsection{Algorithm precision}
\begin{figure}
\begin{center}
\includegraphics[width=\linewidth]{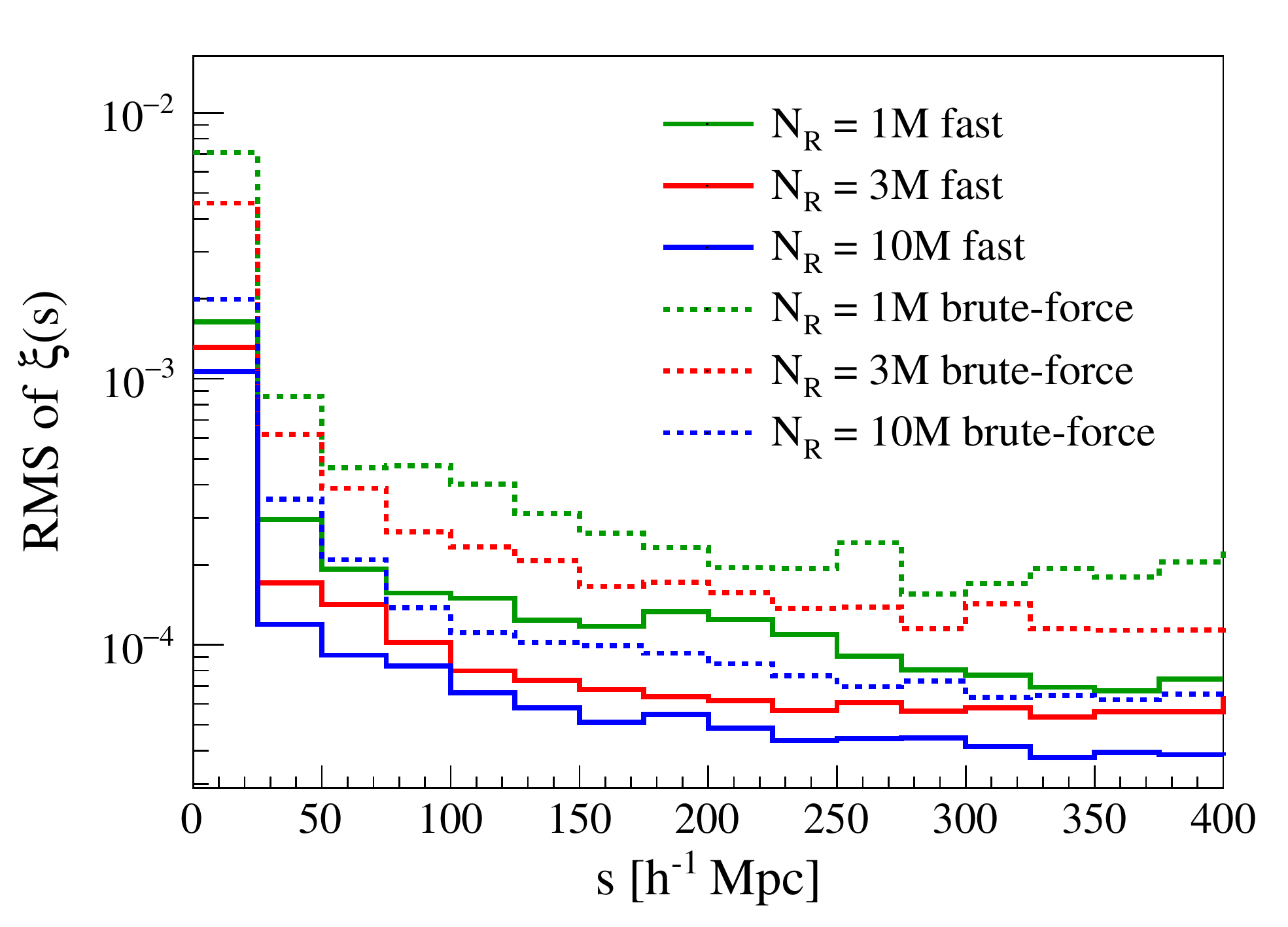}
\caption{Uncertainties in $\hat{\xi}(s)$ as a function $s$ and the size of the
random catalog. The uncertainties are calculated as the RMS of $\hat{\xi}(s)$
obtained with 20 catalogs generated for each of three catalog sizes, for both
methods of computing $\hat{\xi}(s)$.
\label{fig:calib1}}
\end{center}
\end{figure}

To check the precision of the fast algorithm we generate 20 realizations of random catalogs with 3.5M galaxies each using the procedure described above. We estimate the root mean square (RMS) of
$\hat{\xi}(s)$ using the 20 random catalogs and present it in
Fig.~\ref{fig:calib1}. Note that though the bin size used in computation is small as specified above, the result is plotted with a much larger bin size, simply for visual purposes. 
In both the brute-force and fast algorithms, the
uncertainties in $\hat{\xi}(s)$ decrease as the size of the random catalog
increases, but the uncertainties in the results based on the fast algorithm are smaller than
those in the respective brute-force results, in agreement with the
qualitative argument presented in Section~\ref{sec:proof}.

\begin{figure}
\begin{center}
 	\includegraphics[width=\linewidth]{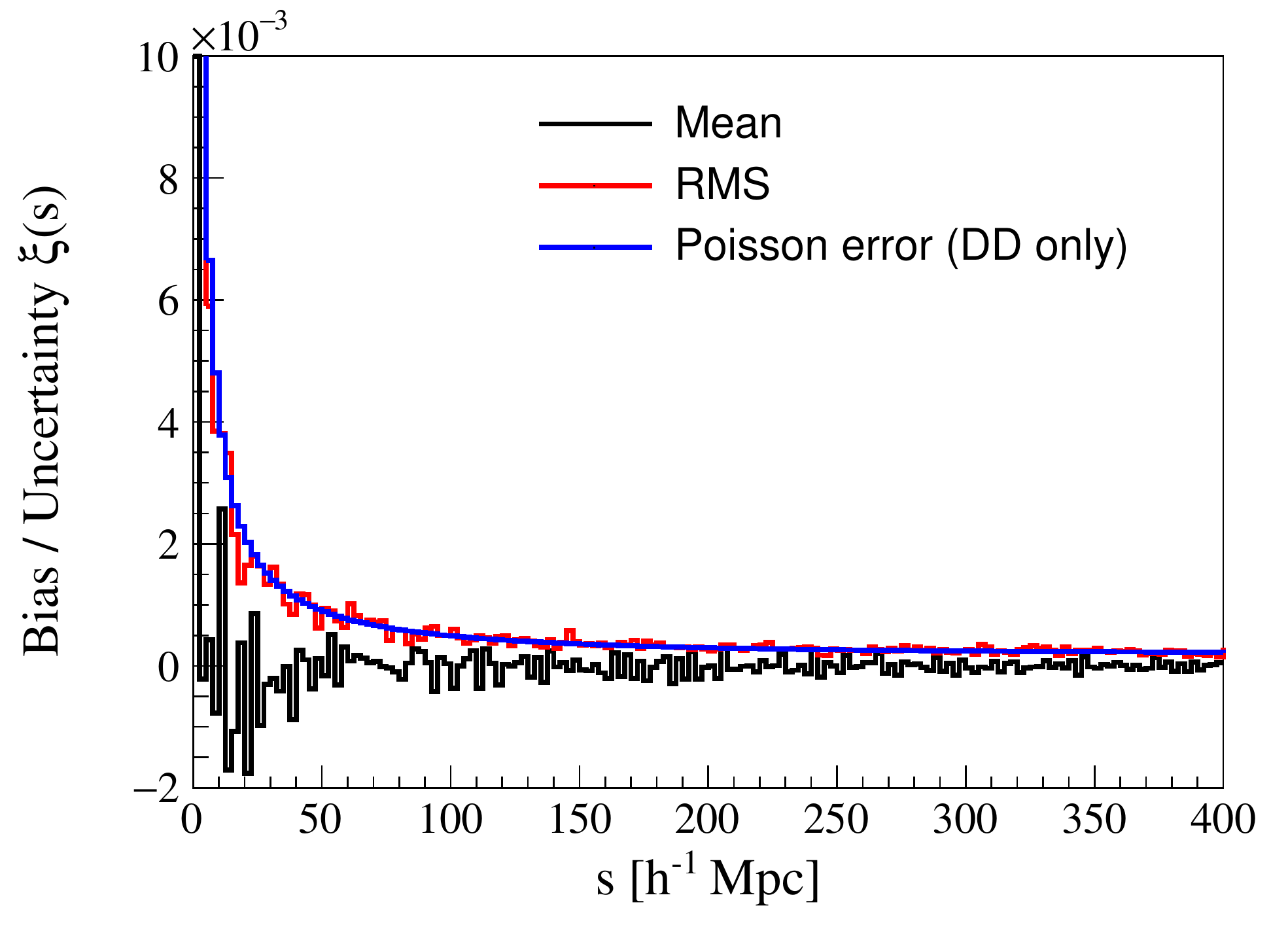}
\caption{Mean and RMS uncertainty in $\hat{\xi}(s)$ of 20 uniformly-generated
mock data sets with 200k galaxies per data set.
\label{fig:calib2}}
\end{center}
\end{figure}

To check the fast algorithm for bias, we compute $\hat{\xi}(s)$ for 20 mock
catalogs representing $D$ with 200k galaxies Poisson-distributed within the
survey volume. These mock catalogs are produced using the same procedure used to generate the random
catalogs. We find no significant bias; that is, the mean of $\hat{\xi}(s)$ is centered at zero as expected for a truly random distribution, and
its RMS obeys a Poisson distribution (Fig.~\ref{fig:calib2}). Note that the Poisson error is dominated by $DD$, since the $DR$ and $RR$ are much larger and hence contribute smaller relative errors.
\subsection{Performance on data with BAO signal}
\begin{figure}
\begin{center}
\includegraphics[width=\linewidth]{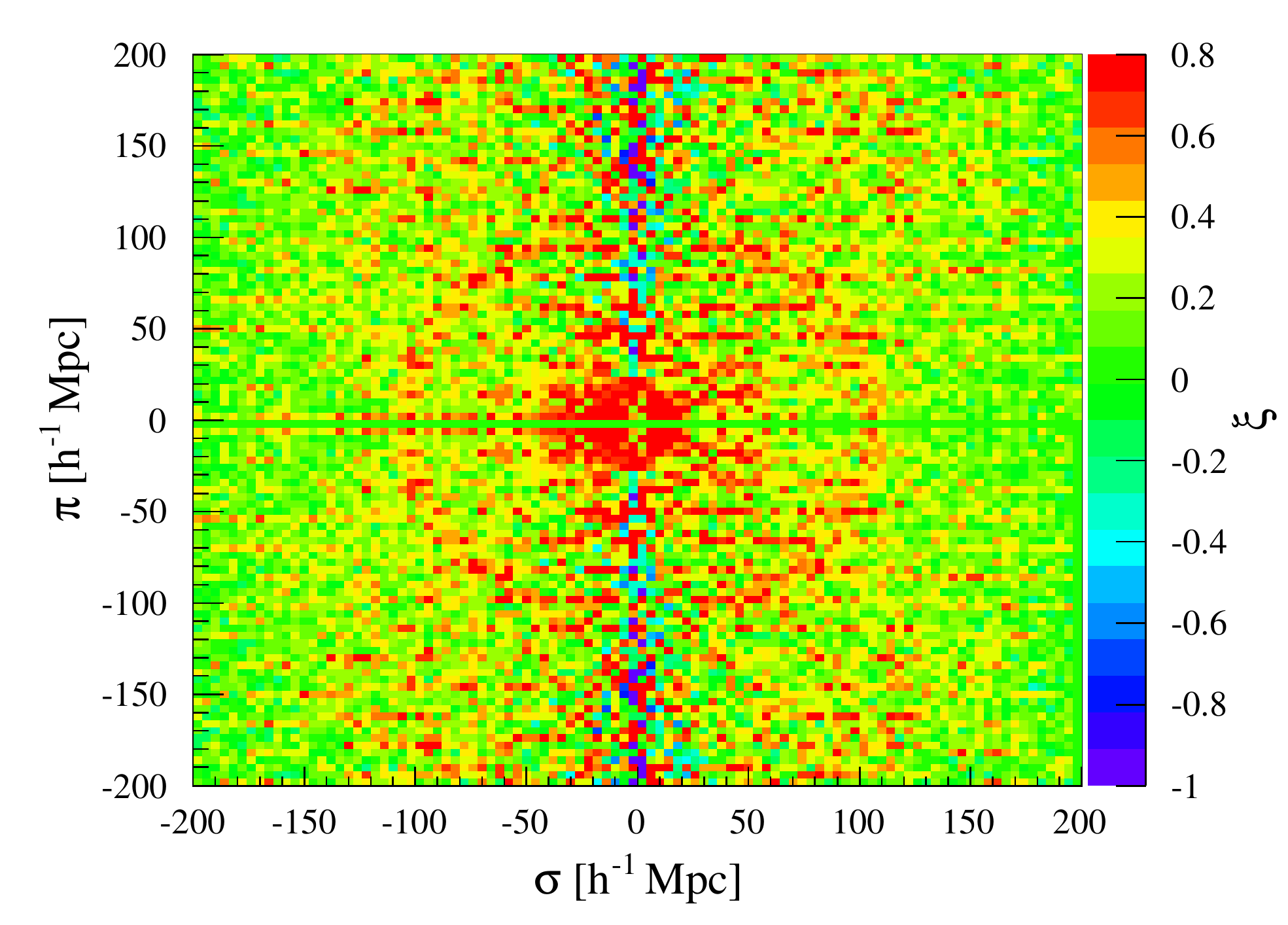}
\caption{ $\hat{\xi}(\sigma,\pi)$ based on a 
mock data set with embedded BAO signal with the radius of $105h^{-1}$~Mpc.  The BAO signal is clearly seen as a circle in the 2D distribution. 
\label{fig:sigma_pi_2D}}
\end{center}
\end{figure}

To demonstrate the performance of the algorithm for the anisotropic case we present the result on a mock dataset with a strong signal embedded (Fig.~\ref{fig:sigma_pi_2D}). The signal was generated by adding spatially correlated galaxy pairs on top of a uniform background. The galaxy pair separation distance is distributed according to a Gaussian of mean $105h^{-1}$~Mpc and a standard deviation of $5h^{-1}$~Mpc. To avoid biasing the $z$ distribution of galaxies in the mock sample, after signal addition some of the galaxies are removed to preserve the original distribution in $z$.

\begin{figure}
  \begin{center}
  \includegraphics[width=\linewidth]{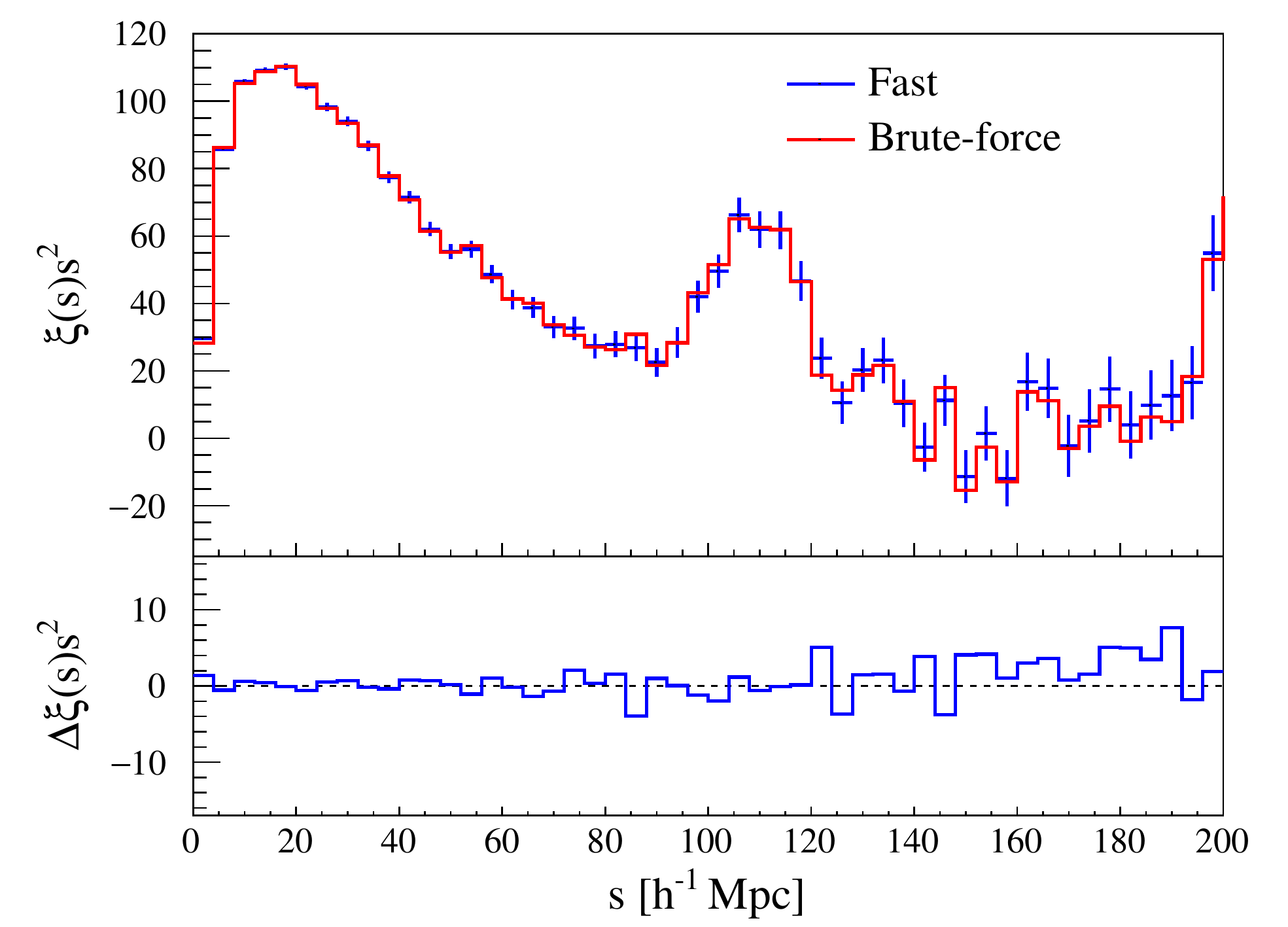}
  \end{center}
  \caption{{\sl Top}:
  $\hat{\xi}(s)\times s^2$ for SDSS-III DR9 BOSS data catalog using the fast
  algorithm suggested in this paper (blue) compared  to brute-force calculation
  (red).  The error bars represent Poisson uncertainties in $DD$.  {\sl
  Bottom}: the difference between the fast and brute-force estimates of
  $\hat{\xi}(s)\times s^2$.}
  \label{fig:data}
\end{figure}

Finally, the algorithm is also applied to the SDSS-III DR9 BOSS data catalog.
The results obtained from the fast algorithm and the brute-force algorithm
and their difference are presented in Fig.~\ref{fig:data}. The BAO peak is
clearly visible in both, and the two distributions are consistent. 
\section{Conclusion}
\label{sec:conclusion}
We have presented a new computational method of the galaxy 2pcf that replaces
summation over all possible galaxy pairs with a numeric integration of the probability map.
The method provides a significant reduction in the calculation time and
improves the precision of the calculation. Moreover, the computationally intensive histogramming part of the calculation is independent 
of the choice of cosmological parameters.  The output of the histogramming stage is used at the fast integration stage, where the cosmological parameters need to be defined to compute the cosmological distances. The integration stage can be repeated for a different set of parameters without redoing the histogramming stage allowing for a fast probe of a larger parameter space.

In the future, the method could be
used for a fast evaluation of the galaxy correlations in large spectroscopic
surveys. In this case, the generation of large-size random catalogs can be
replaced by weighted probability maps determined by observational conditions.
\section*{Acknowledgments}
We thank L. Samuchia, E. Blackman and B. Betchart for useful discussions.  The
authors acknowledge the support from the Department of Energy under the grant
DE-SC0008475..0
\bibliographystyle{mnras}
\bibliography{cosmology-method}

\bsp  
\label{lastpage}
\end{document}